# Teaching AI, Ethics, Law and Policy

**Asher Wilk**[†]
Tel Aviv University
Tel Aviv Israel

## ABSTRACT

The cyberspace and development of intelligent systems using Artificial Intelligence (AI) creates new challenges to computer professionals, data scientists, regulators and policy makers. For example, self-driving cars raise new technical, ethical, legal and public policy issues. This paper proposes a course named **Computers, Ethics, Law, and Public Policy**, and suggests a curriculum for such a course. This paper presents ethical, legal, and public policy issues relevant to building and using intelligent systems.

## 1 Introduction

Recently robots and intelligent systems are equipped with artificial intelligence (AI) and many more will be in the near future. Firms aim to design intelligent systems capable of making their own decisions (autonomous systems). Such systems will need to include moral components (programs) that guide them. For instance, to decide whether a self-driving car should be instructed to swerve to avoid hitting a solid obstacle so as to protect its own occupants even if such a move will lead to hitting a car in another lane. This is an ethical dilemma that reminds us of the trolley problem [82, 83, 64] (e.g., a trolley coming down a track and a person at a switch must choose whether to let the trolley follow its course and kill five people or to redirect it to another track and kill just one).

This paper suggests what should be taught in a course named **Computers, Ethics, Law, and Public Policy**, that is intended for persons involved with AI, new technologies, computer science, information science, and engineering. It presents teaching strategies and suggests teaching ethics and law using examples and case studies demonstrating ethical and legal decision-making. Nowadays, education should not only be about obtaining knowledge, but on addressing critical thinking and decision-making. This paper is based on my experiences in teaching ethics and law to computer science students and to those studying international policy.

**Computer Ethics Education** was not common several years ago, but recently due to the advancements in AI, ethics gained popularity. In 2018 the ACM updated its Code of Ethics and Professional Conduct [66] to respond to the changes in the computing profession since 1992. In the 90s, the code focused on *professionalism* and *quality*. In the new code, the public good is a major concern, and it addresses AI and states: "*Extraordinary care should be taken to identify and mitigate potential risks in machine learning systems*". Ethical issues created by new technologies (e.g., unemployment caused by self-driving vehicles) should not just be the focus of computer professionals - they should concern everybody.

Is *AI Ethics* a new field of study requiring new ethical thinking? Is it different from computer ethics? According to [101], "AI ethics is a sub-field of applied ethics and technology, and focuses on the ethical issues raised by the design, development, implementation and use of AI". Are our current norms sufficient to ensure AI is used and developed responsibly? Do professionals involved with AI need to study several additional ethical disciplines (e.g., business ethics, engineering ethics, and cyber ethics). The above disciplines include many common subjects (e.g., responsibility and accountability) and therefore ethical courses can be structured to address the common subjects in addition to unique subjects.

<u>Structure of this paper:</u> **Section 2** describes the promise and danger of AI. **Section 3** addresses issues relevant to algorithmic decision-making, and ethical and legal problems of autonomous systems, cars and weapons. **Section 4** is about Social Media, Fake News and Journalism. **Section 5** deals with Big Data and Privacy. **Section 6** describes justifications and initiatives for ethics and legal education. **Section 7** on "Law: What to Teach?" suggests what legal subjects to teach, in relation to ethics, such as human rights, privacy and freedom of speech. **Section 8** on "Ethics: What to Teach?" suggests teaching ethical theories, code of ethics, and ethical dilemmas. **Section 9** discusses regulation. **Section 10** deals with programming ethics and laws. **Section 11** concludes with a summary.

## 2   AI: The Promise and the Danger

For many years, the field of AI had limited success. It took many years to develop good expert systems, but they had limited capabilities and addressed specific areas. However, since 2012 dramatic changes occurred because of advancements in Machine Learning (ML) and in particular Deep Learning (DL). The developments in ML allowed rapid progress in various domains including speech recognition and image classification. ML technologies have many beneficial applications, ranging from machine translation to medical image analysis.

AI can also be used maliciously (e.g., cyber-attacks enabled by use of AI to be effective and difficult to attribute, creating targeted propaganda, and manipulating videos). Stanford researchers recently created a facial recognition analysis program that can detect sexual orientation [9]. In parts of the world, such an algorithm could pose a threat to safety. The start-up Faception says its ML technology can spot character traits in a person's face. Such applications raise ethical and legal questions related to discrimination (e.g., is an employer allowed to use face recognition software in the recruitment process?). Computer professionals, policy makers, and actually everybody should be aware of ethical and societal implications of computer and AI applications.

**Employment**: "As automation, robotics, and AI technologies are advancing rapidly, concerns that new technologies will render labor redundant have intensified" [73]. Will AI substitute or complement humans in the workforce? Automation may create new and complex tasks that favor high-skill workers and this will affect inequality. According to [115], "the biggest harm that AI is likely to do to individuals in the short term is job displacement".

**Ethical and legal issues** researchers are concerned with include: unemployment (e.g., caused by self-driving vehicles), inequality resulting from revenues going to fewer people, machines affecting human behavior and inducing social and cultural changes, how to eliminating AI biases (algorithmic biases)? How do we keep AI safe from attacks? How to prevent AI itself to turn against us? How to avoid the point that machines will be more intelligent than humans are? and what rights should robots have?

**Ethics and Law**: New technologies will help us but they also create new moral and legal problems. For instance, problems related to safety, privacy, fairness, human dignity, intellectual property rights, and professional responsibility. New technologies will require updating laws, for example,

*product liability laws* (most of nowadays product liability laws are based on the impact of technologies invented many years ago). Intelligent systems will affect intellectual property (IP) rights and may require changes to *IP laws*. For example, recently Christie's sold an AI-created artwork painted using an algorithm. This raises questions of who should get the IP rights. "Creative works qualify for copyright protection if they are original, with most definitions of originality requiring a human author" [71]. Such generated work could in theory be deemed free of copyright (i.e., could be freely used by anyone) because a human did not create it. That would be harmful for companies selling such works. Copyright laws will have to be updated to address works generated by computers. If we justify IP rights using Locke's labor theory or Hegel's personality theory than there is a problem justifying IP rights for works generated by computers. If the justification is economical, the question is to whom we want to give incentives.

**Trustworthy AI**: An ethical approach to AI is key to generate user trust. According to [101], "AI should be developed, deployed and used with an 'ethical purpose', grounded in, and reflective of, fundamental rights, societal values and the ethical principles of Beneficence (do good), Non-Maleficence (do no harm), Autonomy of humans, Justice, and Explicability. This is crucial to work towards Trustworthy AI". Explicability includes transparency and accountability. To realize Trustworthy AI, requirements should address data governance, human oversight, non-discrimination, privacy, safety, and transparency.

## 3   Autonomous Systems

The power of AI and ML are often demonstrated by their capabilities to win complex games. AlphaGo Zero is the first computer program to defeat a world champion at the ancient Chinese game of Go. It is of great importance because it learns to play by playing games against itself, starting from a completely random play [47]. Scientists had thought it would take many more years to get to that stage. Scholars nowadays are worried how to ensure that computers do not acquire a mind of their own, threatening human existence. Ray Kurzweil a futurist and inventor popularized the idea of "the singularity" - the moment in the future when men and machines will supposedly converge. In *The Singularity is Near* [114] he describes the Singularity, a likely future different from anything we can imagine. According to Kurzweil the moment at which a computer will exhibit intelligent behavior equivalent to that of a human is not far. A *robot brain* would eventually surpass the human brain in rational decision making. Nobel Laureate Daniel Kahneman called for replacing human decision makers with algorithms whenever possible [72].

**What Artificial Intelligence Can and Can't Do Right Now**: According to AI expert Andrew Ng [115], AI will transform many industries, but presently it is still limited ("far from the sentient robots that science fiction has promised us" [115]). Nowadays ML software requires a huge amount of data. You need to show the system many examples so it can learn from them. Data is therefore a barrier for many businesses to use ML.

Machines equipped with AI now make decisions without direct human input. Soon AI may be able to control machines such as automobiles and weapons. Autonomous weapons and cars will affect human safety, privacy, human dignity, and autonomy. They raise new legal and ethical challenges. For instance, who will be held accountable when fully autonomous systems produce unwanted results.

## 3.1   Ethical Concerns

We are witnessing the rise of AI in every aspect of our society leading to a growing recognition of the need to address ethical issues related to AI. Consider for instance *care autonomous robots* developed to support elderly people living at home. Autonomous means here that the robot carries out tasks without continuous human guidance and assistance. Following are relevant ethical and social issues of concern identified in [107]:

- **Implications on employment:** How robots will affect healthcare workers?
- **Trust and reliability:** Shall we trust care provided by robots?
- **Privacy and data protection:** Which data are collected? Who can access the data? Who owns the data?
- **Safety and avoidance of harm:** Robots should not harm people and be safe to work with.
- **Quality of service**
- **Moral agency:** Robots do not seem to have the capacity of moral reasoning or of dealing with ethically problematic situations.
- **Responsibility:** Assuming the robots cannot be morally responsible – who will be responsible?

## 3.2   Autonomous Weapon Systems

The military interest is to increase the autonomy of weapon systems in order to have greater military capability, minimize risks, and reduce costs. Weapon systems with significant autonomy already exist. However, they are not capable of understanding complex and dynamic situations, and making complex decisions like humans do. There are different views on whether in the future "fully autonomous weapon systems" with AI will be in use. There is recognition of the importance of maintaining human control, although it is not so clear what would constitute 'meaningful human control'. The UK policy is that "operation of weapon systems will always be under human control" [2]. Many countries have not yet developed their policy on this or have not discussed it openly. When discussing the ethical and legal problems of using autonomous weapons, international humanitarian law (IHL) should be addressed. Presently programming a machine to make qualitative judgments requires applying "the IHL rules of distinction, proportionality and precautions in attack" [2].

Two central principles from the Just War tradition are the principle of discrimination (distinction) of military objectives and combatants from non-combatants, and the principle of proportionality of means, where acts of war should not yield damage disproportionate to the ends that justifies their use. There are views that current technology is not capable of acting with proportionality and discrimination, and that it is unlikely to be possible in the near future. However, it could be that in the future autonomous systems will be able to deliver excellent results. As AI in military robots advances, the meaning of warfare may change and new international laws may be needed.

Recently several AI researchers called for a ban on offensive autonomous weapons beyond meaningful human control [14]. A campaign to stop killer robots argued that robots and AI lack moral capability and decisions of significant moral weight ought to remain human [18]. According to present laws, "states, military commanders, manufacturers and programmers may be held accountable for unlawful 'acts' of autonomous weapon systems under a number of distinct legal regimes: state responsibility for violations of IHL and international human rights law; international criminal law; manufacturers or product liability; and corporate criminal liability" [2]. Developing autonomous weapon system raises questions of accountability, responsibility, compliance with IHL rules, ethical challenges, and the human right to dignity.

**AI Race**: The Russian president warned that artificial intelligence offers 'colossal opportunities' as well as dangers. "Artificial intelligence is the future, not only for Russia, but for all humankind ... Whoever becomes the leader in this sphere will become the ruler of the world" [69, 70]. This might lead to an AI race similar to the cold war race. In both cases we have prisoner's dilemma type problems.

## 3.3   Autonomous Vehicles

Self-driving cars will have many impacts on society, economy, mobility, the environment, and more. For example, self-driving cars will reduce fatal road accidents and the number of cars and parking lots. Self-driving cars will optimize traffic flow [82], reduce transportation costs and reduce travel time [38]. Self-driving cars will have to deal with potentially life-threatening situations and make moral decisions. Such decisions can be described as an application of the trolley problem [82, 83, 64]. Trolley-like dilemmas take a principle that seems to be a candidate for a universal norm: "Thou shalt not kill" (one of the Ten Commandments), and provide a realistic situation in which an individual kills another human being. The trolley problem can be regarded as an abstraction of many dilemmas involving AI systems. I recommend teaching the trolley problem while looking at the MIT Media Lab Moral Machine experiments [94].

Lin [82] illustrates the need for ethics in autonomous cars, using the following dilemma. An autonomous car must either swerve left and strike an eight-year old girl, or swerve right and strike an 80-year old grandmother. What should be the ethically correct decision? If you were programming the self-driving car, how would you instruct it to behave if it ever encountered such a case? Either choice may be ethically incorrect. What decisions should be made if the law forbids age discrimination? How should a code of ethics that prohibits discrimination based on race, gender, disability, or age be treated? Autonomous driving requires engineers to not only research and develop technology but also to consider ethics and law.

Since automated driving will not be able to avoid all accidents on roads, there is a need to devise laws and rules for what to do in various cases. Some decisions are more than just an application of laws and "require a sense of ethics, and this is a notoriously difficult capability to reduce into algorithms for a computer to follow" [82].

**Liability**: Autonomous Vehicles (AV) will eventually be involved in a situation causing property damage, serious injury, or death to someone. Product liability issues will arise when a vehicle is under the control of the automated system. Manufacturers may be responsible for software updates they distribute but what will be in cases of unauthorized software modifications (e.g., cyber attacks, data security breaches).

Presently, liability in the case of an accident is a controversial topic, in which ethics and economic interests play a role. According to the U.S. National Highway Traffic Safety Administration, states should consider how to allocate liability among owners, operators, passengers, manufacturers, and others when a crash occurs. According to the German Ethics Commission for Automated and Connected Driving, the protection of individuals takes precedence over all other utilitarian considerations. "The licensing of automated systems is not justifiable unless it promises to produce at least a diminution in harm compared with human driving" [11].

Anytime a driverless car makes a decision, it has to make a trade-off between safety and usefulness in a way that is accepted by society. Car manufacturers expect regulators to determine what reasonable decision-making is.

## 3.4 Algorithmic Decision Making

Automated decision-making algorithms are now in use and they have the potential for significant societal impact. For instance, courts use ML algorithms (e.g., the COMPAS risk assessment tool) to establish individual risk profiles and to predict the likelihood that a particular individual will re-offend [85]. "The investigative news organization ProPublica claimed that COMPAS is biased against black defendants" [22]. The company that created the tool released a report questioning ProPublica's analysis.

**Bias problems** of AI software has been reported [108, 22, 85, 109], for example, Amazon developed a recruiting tool for identifying software engineers it might want to hire - the system discriminated against women, and the company abandoned it [108]. Bias can result from cognitive biases of programmers or unrepresentative datasets used for training. The algorithm can be unbiased, but if there is bias in the data used to make the decision, the decision itself may be biased. The IEEE Global Initiative on Ethics for Autonomous and Intelligence Systems [58] aim is to improve fairness of algorithmic decision-making systems.

**Fairness** is very important in algorithmics, but to define fairness and to design fair algorithms is difficult. Fairness can be a data problem - learning algorithms that will learn from biased data will create biased results. Data should be unbiased, but in some cases, it is complicated to determine if data is biased or not.

**The black box problem**: AI systems are often viewed as being 'black boxes' [39] that are complex and difficult to explain. People are less likely to trust machines whose inner workings they do not understand. The black box problem makes it difficult for regulatory bodies to determine whether a particular system processes data fairly and securely. Scholars suggested introducing ethical considerations into algorithms, opening algorithmic black boxes [5, 39], regulating algorithms, and **Fairness**, **Accountability**, and **Transparency** (FAT) in algorithmic decision-making.

**Transparency** facilitates trust in Autonomous Systems. In accident investigation scenarios, transparency helps in diagnosing the causes of errors [37]. International efforts regarding transparency exist (e.g., developing the IEEE P7001 standard on Transparency in Autonomous Systems). Making code and data public may not help a user to understand an algorithm. Therefore, scholars suggested monitoring the outputs of codes to ensure they do not discriminate or cause harm [62]. Another suggestion is to ensure that algorithmic decisions are explainable.

**Explainability**: "The idea of a 'right to explanation' of algorithmic decisions is debated in Europe. That right would entitle individuals to obtain an explanation if an algorithm decides about them (e.g., refusal of loan application)" [62]. According to [13], "without being able to explain decisions taken by autonomous systems, it is difficult to justify them: it would seem inconceivable to accept what cannot be justified in areas as crucial to the life of an individual as access to credit, employment, accommodation, justice and health". According to [37], explainability is at present "virtually impossible with opaque control techniques, such as artificial neural networks". Most machine-learning systems perform "statistical-correlation". They cannot explain their decision. According to Hinton, "People have no idea how they do that … Neural nets have a similar problem", and you should regulate AI algorithms based on how they perform [61]. Many drugs receive regulatory approval, even though no one knows exactly how they work.

**Responsibility** is very important, because in spite of our best efforts, things will go wrong at times. Who should be responsible for decisions made by an algorithm? Who will be responsible to take any necessary actions? Developers should take steps to check and validate the accuracy of the algorithm and the data it uses.

**Verifiability:** Bremner et al. [37] argue that autonomous robots should have an ethical control layer aiming to be proactive, transparent, and verifiable, and that ethical reasoning should be verifiable (i.e., it can be proven to abide by a given code of ethics).

## 4   Social Media, Fake News and Journalism

**Targeted advertising**: Web sites and media influence us and affect our minds. "Social media are systematically exploited to manipulate and alter public opinion" [21]. Targeted social media advertising based on user profiling has emerged as an effective way of reaching individuals. In the case of political advertising, this may present a democratic and ethical challenge, and raise policy and regulation questions.

**"Fake news"**: According to [24], "There is worldwide concern over false news and the possibility that it can influence political, economic, and social well-being" and "Falsehood diffused significantly farther, faster, deeper, and more broadly than the truth in all categories of information". To ensure that false content is not amplified across platforms, scholars claim that there is a need for some

"regular auditing of what the platforms are doing" [26] and "a new system of safeguards" [26].

**Regulation:** Bots (short for 'Chabot') were a big problem during the 2016 U.S. elections, since they were used to influence voters. A 'Bot' is an automated online account where all or substantially all of the actions or posts of that account are not the result of a person action. Effective July 1, 2019, it will be illegal for bots interacting with California consumers to pretend they are human if they are trying to sell goods, services, or to influence a vote in an election. Violators could face fines under state statutes related to unfair competition. This is a legal step in fighting "fake news".

**Social Media Ethics and Regulation**: Social media plays a central role in modern communication. However, social media tools can be abused (e.g., disinformation in the context of political propaganda, extremist groups using social media for radical propaganda and recruitment efforts). Would restricting online speech or imposing new obligations on digital platforms effectively reduce fake news? "Appointing online platforms as 'ministries of truth' to decide what content is desirable and steer users to appropriate channels would be both futile and undemocratic" [103]. Instead of regulation, maybe we should use education. To fight fake news Finland adopted a curriculum and tools to equip elementary and high school students with skills to spot disinformation. Schools are teaching media literacy, fact checking and critical thinking [102].

**Robot journalism**: Nowadays, news organizations such as AP, Reuters, and others are generating thousands of automated stories a month, a phenomenon called "robot journalism". Presently, most uses of robot journalism have been for formulaic situations: stock market summaries and sports stories. Fully automated journalism is going to be very limited for quite a time. However, news companies will be using automatic news writing more and more on challenging subjects. This raises new ethical issues, for instance, the accuracy of the data, and the legal rights to the data. Will the fact that a story was automatically produced be disclosed? Will a human editor check every story before it goes out? Can we program ethical concerns into robot journalism algorithms?

## 5   Big Data and Privacy

Corporations and governments are collecting, analyzing, and sharing detailed information about individuals over long periods. Novel methods for big data analysis are yielding deeper understandings of individuals' characteristics, behavior and relationships. These developments can advance science, benefit individuals and our society, but they are also "privacy risks that multiply as large quantities of personal data are collected over longer periods of time" [25]. ML operating on big data produces results that can violate the privacy of individuals, and can have discriminatory outcomes. "Existing regulatory requirements and privacy practices in common use are not sufficient to address the risks associated with long-term, large-scale data activities" [25]. Notice and consent or de-identification are not enough [25].

## 5.1   Internet of Things (IOT) and Ethics

"The IoT's networked communication will radically alter the way that we interact with technologies" [18]. Devices connected to the IoT collect vast amounts of data and that data can be analyzed and shared. Many devices connected to the IoT have limited or no effective security. This raises concerns of privacy and physical safety.

According to [19], AI based on big data and combined with IoT might eventually govern core functions of society and it is necessary to apply the principles of **rule of law**, **democracy** and **human rights** in AI. With IoT, trust is a major issue. When everything is connected to the Internet, it is hard to trust that information is only shared with those declared. Scholars suggested that people should know about their rights, and that they should be able to control their information and accesses. With IoT, "Informed consent, privacy, information security, physical safety, and trust is foundational" [18].

## 5.2   Privacy: What to Teach?

The Computer Science Curricula 2013 (CS2013) [20] recommends teaching philosophical foundations of privacy rights, legal foundations of privacy protection, privacy implications of wide spread data collection and technology-based solutions for privacy protection [20, 12]. I recommend teaching the approach of Warren and Brandeis [44] on privacy as a "right to be let alone", and the approach of Westin seeing privacy as "the claim of individuals, groups, or institutions to determine for themselves when, how, and to what extent information about them is communicated to others" [45]. I recommend pointing out constitutions and bills of rights addressing privacy, for instance, the European Convention on Human Rights (ECHR), and to address Privacy by Design (PbD) [3]. The idea is to integrate technical privacy principles in system design.

Many laws deal with privacy and different states understand privacy differently. The EU General Data Protection Regulation (GDPR) [10] is probably the best privacy regulation but it is quite complex. I suggest discussing mainly principles and concepts that are in the GDPR, such as: (1) Lawfulness, fairness and transparency (2) Purpose limitations (3) Data minimization (4) Accuracy (5) Storage limitation (6) Integrity and confidentiality. GDPR includes important rights for data subjects that should be addressed, e.g., the right to be forgotten, the right to object to profiling, the right to data portability, and the right to be informed about the collection and use of personal data.

GDPR uses traditional data governance ideas of effective notice, informed consent, and restrictions on the purposes for which data may be used (e.g., 'consent' of the data subject means any freely given, specific, informed and unambiguous indication of the data subject's wishes). Even if regulation puts an emphasis on providing information using clear and plain language, in practice there are privacy policies that are long and hard to read, and an average user may not read them or will find them difficult to understand and correctly interpret. Another problem with 'consent' is that the user is not informed or aware about information generated by AI systems using the data provided by the user (e.g., The "Target" case illustrates the problem [113]).

## 6   Ethics and Law Education

The curricula of many engineering schools include humanities and social science subjects to broaden the scope of education of students. Many firms realized that a company has to be concerned with **corporate social responsibility (CSR)** and business schools are teaching Business Ethics and CSR.

New technologies present new ethical, legal, societal, and public policy issues. Students who will become computer professionals or policy makers should be able to practice their profession in a way that integrates legal and ethical concerns into their work. Recommendations of ethics and law studies to computer students appear in various publications [12, 15, 20, 77]. According to CS2013 [20], graduates should recognize the social, legal, ethical, and cultural issues inherent in the discipline of computing.

## 6.1   Why to Teach Ethics?

**Ethics and Morality**: This paper will not distinguish between morality and ethics. "At its simplest, ethics is a system of moral principles" [81]. Ethics provides us with a moral map that we can use to find our way through difficult issues. "Our concepts of ethics have been derived from religions, philosophies and cultures" [81]. Ethics tells us how to behave in the absence of directions in the law. Both ethics and law are a source of information regarding norms and both help people in decision-making. Making an ethical decision can be a complex task that involves knowing the relevant field (profession) while considering other fields such as law, psychology, sociology, and politics. Unethical behavior is among the greatest personal and societal challenges of our time [100]. The media highlight sensational cases (e.g., British Petroleum's failure to take proper safety precautions in advance of the 2010 oil spill), but "unethical behaviors are committed by 'ordinary' people who value their morality highly but cut corners when faced with an opportunity to gain from dishonest behavior" [100].

After the Enron case and the Great Recession of 2008, business schools viewed business ethics as important to teach. A decade later, can we say that ethic studies improved ethical behaviour? Friedman and Gerstein [88] ask if we are wasting our time teaching business ethics, and describe several ethical lapses at numerous major firms. "Organizational leaders and boards may be paying lip service to the importance of integrity and ethics but are not practicing what they preach" [88].

It has been claimed that the potential of traditional approaches for teaching ethics to transform human behaviour is generally limited [93]. I believe that studying ethics will have a positive influence on some students and they will influence others.

Following are justifications to teach ethics: (1) People perform moral evaluations regularly. Studying ethics can improve their decision-making and improve how they live [1], enrich their lives and the lives of those around them. (2) Ethics provides tools that help make better decisions for the benefit of a person and society ("ethical theory introduces new critical tools for analysis" [97]). (3) Some professions, such as medicine and the law, have traditional codes of ethics, and ethics is part of the professional education. This should be the case also for AI and computer professionals. (4) "Being ethical is simply good business" [55]. (5) Employers are not looking only for technical skills but they prefer to hire persons with good soft skills that include work ethics and responsibility. (6) "Ethics has become a major consideration for young people in their selection of work and career" [55] ("Employees are increasingly challenging technology companies on their ethical choices" [95], Generation Z expect businesses to "practice what they preach when they address marketing issues and work ethics" [112].) (7) Many cases of unethical behaviour suggest that ethics education is important. (8) Law develops slower than new technologies and may not provide solutions to ethical dilemmas. (9) To demonstrate that acts can be legal but not ethical. (10) The public is expecting technology companies to have higher responsibility to the public good.

**Ethics is important for AI developers** since people developing AI technologies have a higher responsibility, as the uncertainty of the methods and applications of machine learning could lead to public distrust. According to [104], "Decisions about the responsible design of artificial intelligence (AI) are often made by engineers with little training in the complex ethical considerations at play". "In view of the magnitude of risk and the central role that AI will have … those responsible must be taught from the beginning how to design for healthy outcomes" [104]. Academic institutions are launching new courses on computer ethics and cyberethics. According to Tavani [40], cyberethics is the study of moral, legal, and social issues involving cyber technology. Cyber technology comprises computing and communication systems. Harvard and MIT are offering a course on AI ethics. Some are integrating ethics across their computing curricula. "The idea is to train the next generation of technologists and policymakers to consider the ramifications of innovations - like autonomous weapons or self-driving cars - before those products go on sale" [16]. Harvard university professor Grosz initiated cooperation between Computer Science and Philosophy aimed at integrating ethics based conversations into existing Computer Science courses ("Embedded EthiCS" initiative).

The course "The Ethics and Governance of Artificial Intelligence" [17] includes concepts such as: Governance, Explainability, Accountability, Transparency, Discrimination, Fairness, Algorithmic Bias, Autonomy, Agency, and Liability, which are important to teach. According to WIPO [34], legal and ethical concerns linked to AI "include the transparency, verifiability and accountability of AI, the right to privacy, the right to equal treatment and avoidance of bias, and the mitigation of negative impacts on employment".

**The ethics of data** focuses on ethical problems posed by the collection and analysis of large datasets. Recently, scholars advocate that ethics be part of the data science curriculum [46, 59]. Data science ethics should address the quality of data and evidence, data accuracy and validity. Data scientists need to address security, privacy and anonymity of data.

## 6.2   Why to Teach Law?

Basic legal knowledge is important for many reasons, including: (1) Legal cases provide good examples on ethical decision-making. (2) It helps to understand the limitations of law and the need to act responsibly. (3) Basic legal education enables computer professionals to interact with law professionals and regulators. (4) Certain areas of law are relevant to computer professionals (e.g., privacy and intellectual property). (5) It helps to understand the relationship between law and ethics. (6) Law is integral to understanding responsibility (e.g., to understand corporate responsibility [92]). (7) Laws reflect and promote societies values (e.g., the European Union GDPR reveals that Europeans value individual privacy and the US Constitution's First Amendment reveals that Americans value free speech).

**Law**: The important goals of law are to maintain order in our social life and to help resolve conflicts. For example, law aims to prevent undesirable behavior using criminal law and tort law, and to create ways to conduct business by using contract law. Law is also a way to create norms. Law influences social norms since people regard law as a source of information that helps them to decide how to behave. Law varies in its content from place to place and from time to time. After World War II, the world changed its attitude to human rights. For example, rights like autonomy, privacy, and freedom of speech became more central. Law protects liberties and rights.

**Law and Ethics**: The idea that what is 'moral' may or may not be also 'legal' is one of the subjects discussed in legal philosophy ('jurisprudence'). Should law be moral? The theory of natural law sees a powerful connection between morality and law, while according to the theory of legal positivism law need not be moral to be a law. The law should be followed simply because it is the law. After World War II, there were claims that there can be a conflict between the obligation to obey the law and the obligation to act morally. Should one consider as "law" all the rules that were formally valid by the Nazi legal system, regardless of their amoral content? In a democratic society where laws are generally taken to be legitimate, something being legal is usually also moral. Ethics can help in the interpretation of laws, in challenging existing laws and in forming new laws.

## 7   Law: What to Teach?

For students to engage fully in the study of any subject they need to find it interesting. There are various alternatives of what subjects of law to teach. Security professionals might be more interested in cyber war and crime, while computer professionals might be more interested in privacy and intellectual property (IP). CS2013 [20] suggests teaching Intellectual Property, Privacy and Civil Liberties, and Professional Ethics. The Computer Engineering Curricula 2016 [15] recommends teaching computer engineers contract law, intellectual property, privacy and secrecy ("privacy and secrecy are fundamental to computing" [15]). Quinn in "Ethics for the Information Age" [28], besides dealing with ethics, has chapters on Intellectual Property, Information Privacy, Privacy and the Government, and Computer Reliability. Spinello in [27] analyzes moral dilemmas and social problems arising in cyberspace, and looks at content control, free speech, intellectual property, privacy and security, and social networking.

In courses I taught on ethics and law, I choose (1) to use case examples that will interest the students. (2) to concentrate on principles. (3) to introduce students to law fundamentals. (4) to use legal case examples that refer to ethical dilemmas.

**Legal subjects related to ethics** include subjects related to human rights and civil liberties, fundamental freedoms,

privacy, the need to balance rights (e.g., freedom of speech vs. defamation) and proportionality (rights are not absolute).

**Law fundamentals** should cover basic legal rights (civil rights) and responsibilities, as well as the difference between criminal law and civil law, explain that law is territorial (it is determined by geographical boundaries) and the problems arising from law being local (national or domestic) while cyber and technology are global. Law fundamentals should include sources of law (common law, statutory law and constitutional law), the common law, and the civil law traditions. I recommend addressing difficulties in legislation, problems of law interpretation, and international law (developed through the adoption of customs and the signing of treaties) including human rights law.

## 8   Ethics: What and How to Teach?

Nowadays ethics is being concerned mainly with what is morally good or bad, right or wrong. Ethical rules and principles differ from one culture or society to another, and have changed from one generation to the next. Therefore, there is place to address various schools and philosophies of ethics. Having a basic understanding of the major ethical theories will help us in decision-making.

In the western philosophy, the following approaches are well known, and I recommend addressing them: (1) **Ethics of consequences** including **utilitarianism** proposed by Jeremy Bentham and John Stuart Mill [51], act utilitarianism and rule utilitarianism. (2) **Ethics of Duty (Deontological ethics),** including formulations by Immanuel Kant [50]. (3) **Virtue Ethics** focuses on virtues (behaviors) that will allow a person to achieve well being. Most known is the ethics of Aristotle [52] that asks what is the character or personality of an ethical person. (4) **Justice** as in ideas of John Rawls [53, 76], and **Social Contract Ethical Perspectives**. (5) **Principlism**, that focuses on the principles of autonomy, beneficence, non-maleficence, and justice [54], and nowadays are the most fundamental ethical principles in medical practice, patient care and treatment.

According to [97] most AI practitioners operate within the ethical framework called "utilitarianism". Goldsmith and Burton [97] first describe utilitarian theory, and then briefly introduce deontology and virtue ethics. There are ethical dilemmas that do not have optimal solutions and different ethical theories will lead to different solutions. "Consider, for example, a utilitarian doctor who harvests one healthy (but lonely and unhappy) person's organs to save five other people. Those five people could go on to experience and create more happiness combined than that one person ever could alone. Although this action is morally problematic, utilitarianism would seem to justify it if it best maximizes utility" [48]. However, there are ethical theories that forbid taking someone's life.

According to [23, 97], ethics should be part of the AI curricula. Scholars claim that ethical thinking is better developed when ethics education is embedded throughout the entire curricula, not just in stand-alone courses.

It is important to look at ethical codes of organizations, firms, and professions. They provide practical guidelines. Mere knowledge of the relevant professional code can be acquired by simple rote learning but the goal of ethics education is far more than that. Ethic codes are important, but to be effective, training, practice, and examples are needed. Lafollette [1] theorizes that ethics is like most everything else that we strive to be good at - it requires practice and effort.

**Ethical Codes**: Nowadays, many organizations have a Code of Ethics and Professional Conduct. "A central reason why companies have ethics codes … is because of court cases" [92]. An ethical code expresses the organization vision and values. It may not provide specific answers to an ethical dilemma, but it provides helpful guidance to resolve difficult ethical situations. Professional organizations also have ethical codes. For example, the Association of Computer Machinery (ACM) has a code "designed to inspire and guide the ethical conduct of all computing professionals" [66]. It is in the interest of a profession to set standards for its members so as to protect the reputation of the profession.

Different disciplines (e.g., medicine, engineering) differ in their code of ethics, but they also have many common values, norms and ideas (e.g., safety, public good, professionalism, responsibility, fairness, human dignity). Different disciplines may emphasize different values (e.g., bioethics may emphasize autonomy and consent; military ethics may emphasize courage, determination, and loyalty; business ethics may emphasize corporate social responsibility, stakeholder theory, whistle blowing, sustainability). Ethical codes including domain-specific ethical codes can never function as a substitute for ethical reasoning itself [101].

A survey [98] found that the most common ethical topics taught to computing students were professional practice issues (64%) and the societal impacts of engineering and technology (62%). Privacy and civil liberties were taught by 48% of computing educators in the U.S. The most used teaching methods (> 60%) were In-class discussion, Case studies, Lecture, and Examples of professional scenarios [98]. Less common (< 30%) were using videos and project based learning.

## 8.1   Values, Principles, and Rules

A code of ethics of a profession needs to include a description of the profession, principles, and rules.

**Principles** are "abstract rules intended to guide decision-makers in making normative judgments in domains like the law, politics, and ethics" [106]. General principles lack specificity and leads to a need for interpretation.

**Rules** provide solutions to very particular situations, while principles are more general and they set standards. For instance, the principle of equal access to justice is part of many legal systems and it is expressed by particular rules.

**Values** are abstract concepts of what a society regards as most worthwhile, for example, the value of equality between human beings. Some view values as more abstract than principles and derive from values principles and from principles specific rules. Some see principles and values as synonyms. From values and principles, rules and laws can be derived, for example, discrimination laws that makes certain activities unlawful and promotes the value of equality between human beings. Values that regard well-being of other people include no suffering (do no harm), autonomy (let people control their own actions, let people make their own plans), and equality. Values that regard well-being include excellence (this relates to virtue ethics) and trust.

When teaching either ethics or law, it is important to point out the difference between principles and rules while demonstrating them. Most of the principles proposed for AI ethics are not specific enough to guide actions. "There is a gap between the abstract rules and the concrete facts to which they apply … Often, there are no rules of finer granularity that elaborate the abstract rules' meaning or explain how they apply in concrete circumstances" [106]. Principles may be in conflict. For example, a truly beneficial application of AI that saves lives might involve using personal data in a way that is in contrast to privacy principles. McLaren [106] shows a conflict between two principles of the National Society of Professional Engineers code, where an engineer's obligation to public safety stands against his obligation to maintain his client's confidence.

**An ethical decision process** involves going from a set of cultural norms, values, and beliefs to principles and from principles to rules, policies and procedures. For example, as part of the human dignity value, conclude the principle of transparency, stating that entities should always take steps to be transparent about their use of data. This principle can be translated into rules and procedures. Principles and rules are used to form a code of ethics. An effective legal or ethical system must build also on principles and cannot be constructed entirely on a rulebook that provides answers to ethical dilemmas or to legal questions. Professional codes of ethics, for instance, the ACM and IEEE, include both rules and principles.

I suggest discussing the four principles of healthcare ethics: Autonomy, Beneficence, Non-maleficence, and Justice [54]. Autonomy is about the right to decide what happen or does not happen to us. It is about an informed consent, which is fundamental in bioethics [49] and appears in privacy laws such as the GDPR [10]. Justice is about fairness and equality. Beneficence is about "Doing good" and acting in the best interests of others. Non-maleficence is about "Do no harm", which is an important principle in the ACM code of ethics. Rendtorff et al. [89] attempted to identify ethical principles related to autonomy, dignity, integrity and vulnerability, which are four important ideas in European bioethics. The research concluded that "the basic ethical principles cannot be understood as universal, everlasting ideas or transcendental truths but they rather function as 'reflective guidelines'".

**Terminology challenges**: Different terms are sometimes used to refer to the same idea and sometimes to different ideas (e.g., 'transparency' and 'explainability'). A term can be interpreted in different ways, for instance, "autonomy" can be interpreted differently in different countries. Autonomy can mean different things in different disciplines and can mean different things in different ethical codes. Autonomy can mean (1) "permission" [90] (2) capacity for creation goals for life (3) capacity for decision and action without outer constraint (4) capacity of political involvement (5) capacity of informed consent, and (6) capacity for privacy [89]. "Problem solving in ethics and law can be characterized as ill-defined" [106] due to lack specific rules, conflicts among abstract rules (principles) and open-textured terms and phrases (e.g., what exactly does it mean to "hold paramount" the safety, health, and welfare of the public?[1]). Therefore, it is difficult to use formal logic in solving ethical and legal problems.

Each profession has particular values and norms. For example, military ethics emphasizes human dignity, loyalty to the country, respecting the rule of law, courage, responsibility, reliability, friendship, dedication, discipline, and personal example [56]. In the late 1940s, Norbert Wiener created the field of "computer ethics". He viewed "The Principle of Freedom", "The Principle of Equality" and "The Principle of Benevolence" as "great principles" upon which society should be built [43].

**AI principles**: The World Commission on the Ethics of Scientific Knowledge and Technology (COMEST) of UNESCO proposed a technology-based ethical framework on robotics ethics [6], that includes principles and values of: (1) human dignity; (2) autonomy; (3) privacy; (4) "Do not harm"; (5) responsibility; (6) beneficence; and (7) justice. Microsoft outlined ten principles to approach AI [7] that clearly state that AI must prevent bias. Google has published principles that will guide Google [8, 105]. Among them are to avoid creating or reinforcing unfair bias, be built and tested for safety, be accountable to people, and incorporate privacy design principles. Zeng [29] lists 27 proposals of AI principles. The proposals address concepts, such as dignity, human rights, equality, fairness, justice, bias, discrimination, transparency, privacy, data protection, security, safety, validation, accountability and responsibility. Such concepts should be addressed in an ethics and law course.

## 8.2  Teaching Strategies

Ethics does not always provide a right answer to moral problems. For many ethical issues there is not a "right" answer. When evaluating ethical thinking one should check if the thinking involves recognizing ethical issues, recognizing ethical responsibilities, suggesting ways to deal with a dilemma, justifying an ethical stand, and proposing a policy regarding an ethical question.

**Discussion**: Teaching ethics should involve discussing ethical dilemmas and how to make ethical decisions. Discussion is important for increasing understanding and for seeing multiple viewpoints. Discussion involves active participation rather than just listening to lectures, and can enhance abilities of students to recognize ethical issues. Some business schools use video and film to present ethical dilemmas followed by discussions (e.g., a film about the Enron case [87]).

**Seminar format:** Rendtorff [91] presents teaching business ethics in a seminar format. "Students have approximately one to two months to prepare their papers (12-15 pages) and they have approximately 10-15 minutes to present their papers in class with up to 20 minutes of discussions following the papers including the commentaries from the opponents" [91].

Beaton [80] suggests several teaching methods including holding a trial, holding a debate, and presentations by a group of students. "The benefit of holding a trial is that students must do research and be prepared not only to defend what they consider to be their cases' strengths, but to also be prepared to tear apart what they think the other team will argue. Thus they get a dual perspective of the issue being addressed" [80].

**Balancing theory and practice**: There is a debate about the optimal balance between the theoretical ethics (philosophy) and practical instruction (cases) [78]. Burton [23] suggests teaching by presenting case studies, discussing ethical issues and analyzing them using utilitarianism, deontological, and virtue theories.

---

[1] "Engineers, in the fulfillment of their professional duties, shall hold paramount the safety, health, and welfare of the public in the performance of their professional duties". National Society of Professional Engineers. 1996.

**Decision Making Methodologies:** The known legal analysis methodology IRAC (Issue, Rule, Application, and Conclusion) can be used also for ethical analysis. It is a problem-solving tool. In legal analysis, the "issue" is a legal question that arises out of the facts presented and the "rule" is law that applies to the legal question (in a common law jurisdiction rule is derived from court case precedent and statutes). The "application" applies the rules to the facts of the issue at hand. The "conclusion" answers the question presented in the "issue".

Mepham [74, 75] suggests a practicable framework for ethical analysis using an "ethical matrix" that includes "a set of relevant prima facie principles, and a list of the agents that have 'interests'" [74, 75]. The list of agents will depend on the nature of the issue to be analyzed. Ethical decision-making can use a methodology similar to the one used by the National Association of Social Workers [30]: (1) Determine the ethical dilemma. (2) Identify the key values and principles involved. (3) Rank the values or principles that are most relevant to the dilemma. (4) Develop an action plan that is consistent with the ethical priorities that have been determined as central to the dilemma. (5) Implement your plan, utilizing the most appropriate practice skills. (6) Evaluate the consequences for those involved.

Another strategy that can be used is Brey's method [41, 40]: (1) Identify a practice or a technological feature that is controversial from a moral perspective. Determine whether there are any specific guidelines (professional codes) that can help resolve the issue. (2) Analyze the ethical issue by clarifying concepts and situating it in a context. (3) Deliberate on the ethical issue by applying one or more ethical theories and justifying the position you reached by using logic and critical thinking.

**Subjects to address**: Brey [41] suggests addressing four key values as starting points for studies in computer ethics: justice, autonomy, democracy and privacy.

Stanford University course CS 181: Computers, Ethics, and Public Policy [63], uses case studies in four principal areas: (1) Risk and professional responsibility (e.g., Therac 25 case). (2) Surveillance and privacy (e.g., NSA Surveillance). (3) Gender, race, and participation (e.g., Employment discrimination). (4) Hacking and information.

Prentice [86] suggests teaching about heuristics and biases that influence decision making (e.g., overconfidence, framing [32], cognitive dissonance, sunk costs, loss aversion). "In many settings people are subject to various heuristics and bias that systematically prevent their decision making from being objectively optimal" [86] and could lead to unethical behavior. "Educating students about these heuristics and biases may help minimize their effects" [86].

In many fields of study, there is information overload (e.g., in medicine) and to add to this, AI applications that generate diagnostics and recommendations. There is a need to educate professionals (e.g., physicians) to make decisions taking into account recommendation systems that use AI and big data, and the possible biases.

**Behavioral Ethics**: It is relatively a new field that seeks to understand how people think and behave when confronted with ethical dilemmas. "Its findings show that people are often influenced, subconsciously, by psychological biases, organizational and social pressures, and situation factors that impact decision making and can lead to unethical action" [99]. There are findings that that "people of good character … may do bad things because they are subject to psychological shortcomings or overwhelmed by social pressures, organizational stresses, and other situational factors" [93]. Research shows that when faced with a tempting situation, people can violate moral rules but feel moral due to self-serving justifications that provide reasons for questionable behaviors and make them appear less unethical [100]. "Many deceptive practices fall in a gray area where it is difficult to identify or establish that they are fraudulent with intent to deceive as defined under law" [111].

# 9   Ethics and Regulation

Scholars "are debating where legal-regulatory frameworks are needed and when, if ever, ethical or technical approaches suffice" [62]. Microsoft has recently publicly called for government regulation around facial recognition technologies ("We live in a nation of laws and the government needs to play an important role in regulating facial recognition technology" [96]). According to Microsoft, self-regulation and responsibility are an inadequate substitute for decision making by the public and its representatives in a democratic republic".

**Formal regulation** involves legislation and regulators that have appropriate powers and resources in order to achieve compliance of regulatees. **Non-formal regulation** includes organizational self-regulation and internal codes of conduct and ethics. But self-regulation is problematic. According to Braithwaite [110], "self-regulation has a formidable history of industry abuse of this privilege". According to Gunningham and Sinclair [116], "'voluntarism' is generally an effective regulatory element only when it exists in combination with 'command-and-control' components".

Although many firms have ethical codes, there are cases where firms acted not only unethically but also illegally. The recent case of Volkswagen regarding compliance with emission control standards highlights the critical importance of oversight, an ethical culture, and a compliance system. In the fallout of Enron, laws like Sarbanes Oxley were formed for the purpose of protecting the public and the business from fraud or errors. Sarbanes-Oxley intended to restore confidence in the securities markets by regulating the accounting profession. According to [111], "laws and regulations may fail to achieve their goals if governments do not enforce them consistently and effectively".

The solution to reducing car accidents was not ethics training for drivers, but public policy and laws addressing the safety of cars, the legal duty to put on seatbelts, the safety of roads, licensing of drivers, drunk-driving laws, and the like. "In creating laws and regulations, the key should be first on prevention of harm if it can be achieved at a reasonable cost rather than focusing on how to deal with the conduct after the fact. For example, preventing traffic accidents through appropriate traffic laws such as speed limits and proper infrastructure is better than relying solely on insurance, fines, prisons, civil litigations and ambulances" [111].

Targeted advertising on the Internet is very profitable and it is unlikely companies will abandon this business model because of ethical reasons. The solution is in regulations such as the GDPR.

The Information Technology (IT) industry has successfully lobbied for decades against any attempt to regulate IT, claiming "regulation stifles innovation". Regulation that uses strict rules can impede innovation, might limit technology, and interfere with the development process. However, regulation is sometimes necessary, especially to ensure safety. According to AI now 2018 [5], "There is an obvious need for accountability and oversight in the industry, and so far the move toward ethics is not meeting this need. This is likely in part due to the market-driven incentives". Arguments that are now being presented against legislation for AI have been presented against legislation for data protection (GDPR) [19]. Those arguments include that "the law is not able to develop as fast as technology" [19], "the law is not precise enough to regulate complex technology" [19], and that the law is "not providing sufficient legal certainty now … and not sufficiently open to provide flexibility for the future" [19]. Scholars claim that such arguments are a way of saying what corporations have always said: we want no obligations by law as with laws we could be held accountable through enforcement. Business has no problem with the fact that any ethics code lacks democratic legitimacy and cannot be enforced. "The law has democratic legitimacy and it can be enforced, even against powerful mega corporations" [19].

Creating new laws sometimes takes many years because of difficulties to understand the problems, priorities, and political disagreements. Regulating intelligent autonomous systems is also difficult because (1) it is difficult to know how future intelligent systems will work, and (2) technology is global, while regulation is usually local. There is a need for global regulation of AI but past efforts at global regulation of other matters indicate that it is very difficult. Since regulation has its problems, it is also important to educate people to be responsible and behave ethically.

## 10   Programming Ethics and Laws

An obstacle to automating ethical decisions is the disagreement among moral philosophers on ethical principles [33]. According to Awad [94], "even if ethicists were to agree on how autonomous vehicles should solve moral dilemmas, their work would be useless if citizens were to disagree with their solution". "Ethical issues vary according to culture, religion, and beliefs" [60]. Human ethical decisions are affected by emotions, social upbringing, maturity, gut instinct, and philosophical views [36] and this is impossible to reduce to fixed codes. Such ability is built up by years of parenting, socialization, and involvement in cultures, societies, and communities.

Scholars suggested that autonomous systems should only be deployed in situations where there is a consensus on the relevant ethical issues [57]. According to [35], when ground truth ethical principles are not available, we must use "an approximation as agreed upon by society". For self-driving cars, surveys such as the Moral Machine can help to come up with a social consensus (e.g., a survey of 2.3 million people worldwide revealed that moral choices are not universal. However, almost universally people preferred to save the lives of many over few and the lives of the young over old [94]).

How self-driving cars should be programmed to handle situations in which harm is likely either to passengers or to others outside the car? A possible strategy, following the ideas in [4], is to use technical ways to avoid a dilemma while obeying legal constraints, and if not possible then deciding based on learning from preferences of voters (e.g., using preference data collected from millions of people through the Moral Machine website). It is questionable if ML can arrive at ethical or legal rules. This is a too difficult task for current ML technology. In addition, humans still want to control what rules should machines use.

In the near future, most likely autonomous system will use rules created by people or rules controlled and verified by people. These rules will be specific for particular situations. Intel developed a model called Responsibility-Sensitive-Safety (RSS) [31]. RSS is designed to achieve three goals: (1) the interpretation of the law should be sound in the sense that it complies with how humans interpret the law (2) the interpretation should lead to a useful driving policy (3) the interpretation should be efficiently verifiable [31]. RSS is constructed by formalizing the following five "common sense" rules: (1) Do not hit someone from behind. (2) Do not cut-in recklessly. (3) Right-of-way is given, not taken. (4) Be careful of areas with limited visibility. (5) If you can avoid an accident without causing another one, you must do it.

Each of these general rules (principles) needs to be programmed for different scenarios, involving mathematical computations. A classical approach will express the goal of the system in the form of a cost function that we aim to maximize or minimize. The cost function operates on a set of inputs. Weights are associated with the inputs. This method actually operates according to act utilitarianism in philosophy. In both engineering and philosophy, the fundamental challenge with such an approach lies in developing an appropriate cost function. Goodall [84] illustrates how cost functions can result in unintended consequences. He presents the example of a vehicle that chooses to hit a motorcyclist with a helmet instead of one without a helmet since the chance of survival is greater.

There are situations that taking a deontological, or rule-based, approach to dilemma situations is more appropriate and there are situations where taking a utilitarian approach has advantages. "Ethical principles are typically stated with varying degrees of vagueness and are hard to translate into precise system and algorithm design" [60].

It was suggested to include ethical modules in software (e.g., an ethical layer [37]) and that they should be transparent and verifiable (e.g., "it should be possible to verify that it respects the values it reasons about, particularly … where safety is concerned" [37]). "Safety-critical systems … require high standards of validation - preferably formal verification. … robotic systems … will also require high standards of validation" [37].

What presently can be done is devising specific rules for specific situations using limited information and techniques involving calculations, probabilities, and formal logic. This far from ethical decisions by humans. The decisions made by algorithms will have to be tested and verified using techniques such as simulation, testing and formal verification [65], taking into consideration many possible real world scenarios.

## 11 Summary

This paper explained the importance of teaching ethics and law fundamentals to computer professionals and decision makers that includes legal aspects, ethical aspects, and professional responsibility. Future jobs for computing graduates will require not only technical knowledge but also ethical and legal awareness. We described relevant topics and challenges computer professionals and decision makers should address. New technologies will require developing ethics and law.

## REFERENCES

[1]   H. Lafollette. 2007. *The Practice of Ethics*. Blackwell Publishing.
[2]   Autonomous weapon systems: Technical, military, legal and humanitarian aspects. Expert meeting, Geneva, Switzerland, 26-28 March 2014.
[3]   A. Cavoukian, 7 Foundational Principles, https://www.ipc.on.ca/wp-content/uploads/Resources/7foundationalprinciples.pdf
[4]   R. Noothigattu, N.S. Gaikwad, E. Awad, S. Dsouza, I. Rahwan, P. Ravikumar, and A.D. Procaccia. 2018. A Voting-Based System for Ethical Decision Making. https://arxiv.org/abs/1709.06692
[5]   AI now report (2017, 2018) AI Now Institute, New York University. https://ainowinstitute.org/
[6]   World Commission on the Ethics of Scientific Knowledge and Technology. 2017. Report of COMEST on robotics ethics SHS/YES/COMEST-10/17/2REV. https://unesdoc.unesco.org/ark:/48223/pf0000253952
[7]   A.P. Chaudhry (2017) Microsoft CEO Satya Nadella rolls out 10 rules that will define Artificial Intelligence, May 5, 2017. http://analyticsindiamag.com/microsoft-ceo-satya-nadella-rolls-10-rules-will-define-artificial-intelligence/
[8]   Sundar Pichai (2018) *AI at Google: our principles*. https://blog.google/technology/ai/ai-principles Published Jun 7, 2018
[9]   Heather Murphy (2017) Why Stanford Researchers Tried to Create a 'Gaydar' Machine. *The New York Times*, 9, October 2017.
[10]  European Union (2018) The EU General Data Protection Regulation. https://eugdpr.org/the-regulation/
[11]  Federal Ministry of Transport and Digital Infrastructure - Ethics Commission (2017) Automated and Connected Driving. Jun. 2017.
[12]  Asher Wilk (2016) Cyber Security Education and Law. *2016 IEEE International Conference on Software Science, Technology and Engineering*. https://ieeexplore.ieee.org/abstract/document/7515415
[13]  C. Villani (2018) For a Meaningful Artificial Intelligence. Towards a French and European strategy. *Villani Report*, 8.
[14]  E. Guizzo and E. Ackerman (2016) Do We Want Robot Warriors to Decide Who Lives or Dies? *IEEE Spectrum*. 31 May 2016. https://spectrum.ieee.org/robotics/military-robots/do-we-want-robot-warriors-to-decide-who-lives-or-dies
[15]  Computer Engineering Curricula 2016: Curriculum Guidelines for Undergraduate Degree Programs in Computer Engineering, Interim Report, Oct 2015. Final Report, December 2016. ACM and IEEE Computer Society.
[16]  N. Singer (2018) Tech's Ethical 'Dark Side': Harvard, Stanford and Others Want to Address It. *The New York Times*. Feb. 12, 2018.
[17]  Joi Ito, Jonathan Zittrain (2018) The Ethics and Governance of Artificial Intelligence (Syllabus). MIT, spring 2018.
[18]  Fritz Allhoff, Adam Henschke (2018) The Internet of Things: Foundational ethical issues. *Internet of Things*, Volume 1–2, 2018, pp. 55–66.
[19]  Paul Nemitz (2018) Constitutional democracy and technology in the age of artificial intelligence. *Philosophical and Engineering Sciences*, 376(2133) https://doi.org/10.1098/rsta.2018.0089
[20]  Joint Task Force on Computing Curricula, Association for Computing Machinery (ACM) and IEEE Computer Society (2013) Computer Science Curricula 2013: Curriculum Guidelines for Undergraduate Degree Programs in Computer Science. ACM, New York, NY, USA.
[21]  E. Ferrara. 2017. Disinformation and Social Bot Operations in the Run Up to the 2017 French Presidential Election.
[22]  Avi Feller, Emma Pierson, Sam Corbett-Davies and Sharad Goel. 2016. "A computer program used for bail and sentencing decisions was labeled biased against blacks. It's actually not that clear", *The Washington post*, October 17, 2016.
[23]  E. Benton, J. Goldsmith, S. Koenig, B. Kuipers, N. Mattei, and T. Walsh. 2017. Ethical Considerations in Artificial Intelligence Courses. arXiv:1701.07769v1.
[24]  Soroush Vosoughi, Deb Roy and Sinan Aral (2018) The spread of true and false news online. *Science* 359 (6380), pp. 1146-1151, March 2018.
[25]  Micah Altman, Alexandra Wood, David R O'Brien, Urs Gasser (2018) Practical approaches to big data privacy over time. *International Data Privacy Law*, Volume 8, Issue 1, pp. 29–51, February 2018. https://doi.org/10.1093/idpl/ipx027

[26] David Lazer, Matthew A. Baum, Yochai Benkler, et al. 2018. The science of fake news. *Science*, 359(6380):1094-1096.

[27] R. Spinello. 2016. Cyberethics: Morality and Law in Cyberspace (6th ed.)

[28] Michael J. Quinn. *Ethics for the Information Age* (7th ed.)

[29] Yi Zeng, E. Lu, C. Huangfu (2018) Linking Artificial Intelligence Principles. https://arxiv.org/abs/1812.04814

[30] https://www.socialworkers.org/pubs/code/

[31] S. Shalev-Shwartz, S. Shammah, A. Shashua (2017) On a Formal Model of Safe and Scalable Self-driving Cars. https://arxiv.org/pdf/1708.06374.pdf

[32] A. Tversky, and D. Kahneman. 1981. The framing of decisions and the psychology of choice. *Science*. 211(4481):453-458, 1981.

[33] B. Williams (1986) Ethics and the Limits of Philosophy. Harvard University Press.

[34] WIPO (2019) WIPO Technology Trends 2019: Artificial Intelligence. Geneva: World Intellectual Property Organization.

[35] C. Dwork, M. Hardt, T. Pitassi, et al. 2012. Fairness through awareness. *Innovations in Theoretical Computer Science*, pp. 214-226.

[36] Neil McBride, Robert R. Hoffman (2016) Bridging the Ethical Gap: From Human Principles to Robot Instructions. *IEEE Intelligent Systems*, 30(5), pp. 76-82, Sept. 2016.

[37] P. Bremner, L.A. Dennis, M. Fisher, and A.F. Winfield. 2019. On Proactive, Transparent, and Verifiable Ethical Reasoning for Robots, *Proc. of the IEEE*, 107(3):541-561, March 2019.

[38] Margarita Martínez-Díaz, Francesc Soriguera, Ignacio Pérez (2019) Autonomous driving: a bird's eye view. *Proc. of the IEEE. IET Intell. Transp. Syst.*, Vol. 13 Issue 4, pp. 563-579.

[39] F. Pasquale (2016) The black box society: the secret algorithms that control money and information. Cambridge, MA: Harvard University Press. 320 pages.

[40] H. T. Tavani (2016) Ethics and Technology: Controversies, Questions, and Strategies for Ethical Computing (5th ed.) Wiley, December 2015.

[41] Philip Brey. 2000. Disclosive computer ethics. *SIGCAS Comput. Soc.* 30, 4 (December 2000), 10-16.

[42] Y.N. Harari (2018) *21 Lessons for the 21st Century* (1st ed.) Spiegel & Grau.

[43] N. Wiener (1950) *The Human Use of Human Beings: Cybernetics and Society*, Boston: Houghton Mifflin, (2nd ed. Revised New York, NY: Doubleday Anchor, 1954.

[44] S.D. Warren and L.D. Brandeis. 1890. The Right to Privacy. *Harvard Law Review*, 4(5): 193-220, December 15, 1890.

[45] Alan F. Westin. 1967. *Privacy and Freedom*.

[46] Jeffrey S. Saltz, Neil I. Dewar, and Robert Heckman. 2018. Key Concepts for a Data Science Ethics Curriculum. In *Proceedings of the 49th ACM Technical Symposium on Computer Science Education* (SIGCSE '18). ACM, New York, NY, USA, 952-957.

[47] D. Silver et al. 2017. Mastering the game of Go without human knowledge, *Nature* 550:354–359. See also David Silver et al. A General Reinforcement Learning Algorithm That Masters Chess, Shogi, and Go Through Self-play, *Science* 362:1140-44, 2018.

[48] C. Grau (2006) There Is No "I" in "Robot": Robots and Utilitarianism. *IEEE Intelligent System*. 21(4), pp. 52-55, July-August 2006.

[49] UNESCO (2006) Universal Declaration on Bioethics and Human Rights. http://unesdoc.unesco.org/images/0014/001461/146180E.pdf

[50] Immanuel Kant (1785) Fundamental Principles of the Metaphysic of Morals. https://www.gutenberg.org/ebooks/5682

[51] https://www.utilitarianism.com/

[52] Aristotle. *Nicomachean Ethics*. Translated by W.D. Ross. http://classics.mit.edu/Aristotle/nicomachaen.html

[53] J. Rawls (1999) *A Theory of Justice*. Harvard University Press. Revised Edition.

[54] Tom L. Beauchamp, James F. Childress (2012) *Principles of Biomedical Ethics*. Oxford University Press.

[55] https://opentextbc.ca/businessethicsopenstax/

[56] CF Code of Values and Ethics. http://www.forces.gc.ca/en/about/code-of-values-and-ethics.page (accessed 21 July 2019)

[57] M. Anderson and S.L. Anderson (2011) *Machine Ethics*. New York: Cambridge University Press.

[58] IEEE Standard Association (2018) *Ethically Aligned Design, version 2.* https://ethicsinaction.ieee.org/

[59] https://www.coursera.org/learn/data-science-ethics

[60] National Science and Technology Council (2016) National Artificial Intelligence Research and Development Strategic Plan. October 2016.

[61] Hessie Jones, "Geoff Hinton Dismissed The Need For Explainable AI: 8 Experts Explain Why He's Wrong". Forbes, 20 Dec. 2018.

[62] C. Cath (2018) Governing artificial intelligence: ethical, legal and technical opportunities and challenges. *Phil. Trans. R. Soc.* A376: 20180080. http://doi.org/10.1098/rsta.2018.0080

[63] Stanford University (2018) CS 181/181W: Computers, Ethics, and Public Policy, Fall 2018. https://stanfordcs181.github.io/

[64] P. Foot. 1967. The problem of abortion and the doctrine of double effect. Oxford Review, 5:5–15, 1967.

[65] Asher Wilk. 1990. On system reliability and verification. *COMPEURO'90: 1990 IEEE International Conference on Computer Systems and Software Engineering*. https://ieeexplore.ieee.org/document/113660

[66] https://www.acm.org/code-of-ethics

[67] https://ethics.acm.org/code-of-ethics/software-engineering-code/

[68] House of Lords Select Committee on Artificial Intelligence (2018) Report of Session 2017–19, AI in the UK: ready, willing and able? (HL paper 100)

[69] James Vincent. 2017. Putin says the nation that leads in AI 'will be the ruler of the world' https://www.theverge.com/2017/9/4/16251226/russia-ai-putin-rule-the-world

[70] T. Simonite. 2017. For Superpowers, Artificial Intelligence Fuels New Global Arms Race. *Wired, 8 September 2017.*

[71] M.A Guadamuz, Artificial intelligence and copyright, WIPO Magazine 5/2017.

[72] D. Kahneman. 2018. NBER Economics of Artificial Intelligence Conference, Toronto, Canada. See Joshua Gans, Digitopoly blog, https://digitopoly.org/2017/09/22/kahneman-on-ai-versus-humans/ and see https://youtu.be/gbj_NsqNe7A

[73] Acemoglu, Daron and Pascual Restrepo. 2018. The Race Between Machine and Man: Implications of Technology for Growth, Factor Shares, and Employment. *American Economic Review* 2018, 108(6):1488–1542. https://doi.org/10.1257/aer.20160696 http://ide.mit.edu/sites/default/files/publications/aer.20160696.pdf

[74] B. Mepham (2006) *Bioethics*, USA: Oxford University Press.

[75] B. Mepham (1996) Ethical analysis of food biotechnologies: an evaluative framework. In: *Food Ethics*. B. Mepham (editor) pp. 101–119 London, Routledge,

[76] J. Rawls (2001) *Justice as Fairness: a restatement*. Belknap Press.

[77] A.R. Bielefeldt, M. Polmear, D. Knight, C. Swan, N. Canney (2018) Education of Electrical Engineering Students about Ethics and Societal Impacts in Courses and Co-curricular Activities. *IEEE Frontiers in Education Conference.*

[78] D.R. Haws. 2001. Ethics Instruction in Engineering Education: A (Mini) Meta-Analysis. *Journal of Engineering Education*, 90(2):223-229.

[79] C. Fiesler (2018) Tech Ethics: A Collection of Syllabi.

[80] Catherine Beaton. 2009. Creative ways to teach ethics and assess learning. In Proceedings of the 39th IEEE international conference on Frontiers in education conference (FIE'09). IEEE Press, 897-900.

[81] http://www.bbc.co.uk/ethics/introduction/intro_1.shtml

[82] P. Lin . 2016. Why Ethics Matters for Autonomous Cars. M. Maurer et al. (eds.) *Autonomous Driving*, DOI 10.1007/978-3-662-48847-8 (eBook)

[83] D. Edmonds. 2014. Would You Kill the Fat Man? The Trolley Problem and What Your Answer Tells Us About Right and Wrong. Princeton University Press, Princeton.

[84] N.J. Goodall. 2014. Machine ethics and automated vehicles. G. Meyer, and S. Beiker (eds.) *Road Vehicle Automation*. Springer.

[85] See State v. Loomis, 881 N.W.2d 749 (Wis. 2016).

[86] Robert Prentice (2004) Teaching Ethics, Heuristics, and Biases. *Journal of Business Ethics Education (JBEE)*, Vol. 1, Issue 1, 57-74.

[87] Enron: The smartest guys in the room (2005) Magnolia Studios.

[88] H.H. Friedman, M. Gerstein (2016) Are We Wasting our Time Teaching Business Ethics? Ethical Lapses Since Enron and the Great Recession. http://ssrn.com/abstract=2839069

[89] J.D. Rendtorff, P. Kemp (2019) Four Ethical Principles in European Bioethics and Biolaw: Autonomy, Dignity, Integrity and Vulnerability. In E. Valdés, J. Lecaros (eds) *Biolaw and Policy in the Twenty-First Century*. International Library of Ethics, Law, and the New Medicine, Vol. 78, Springer, Cham.

[90] T.H. Engelhardt (1996) The foundations of bioethics (New Revised ed.). New York: Oxford University Press.

[91] Jacob Dahl Rendtorff (2015) An Interactive Method for Teaching Business Ethics, Stakeholder Management and Corporate Social Responsibility (CSR). *Journal of Business Ethics Education* 12 Special Issue: 93-106.

[92] Timothy Fort, Stephen Presser (2017) The Legal Environment of Business. West Academic Pub.

[93] Robert Prentice (2014) Teaching Behavioral Ethics. *Journal of Legal Studies Education*, Volume 31, Issue 2, 325–365.

[94]  Edmond Awad et al. 2018. The Moral Machine Experiment. *Nature* 563: 59-64, https://www.nature.com/articles/s41586-018-0637-6

[95]  Keith Kirkpatrick. 2019. Ethics in technology jobs. Commun. ACM 62, 6 (May 2019)

[96]  Brad Smith. 2018. Facial recognition technology: the need for public regulation and corporate responsibility. *Microsoft Issues,* July 13, 2018. https://blogs.microsoft.com/on-the-issues/2018/07/13/facial-recognition-technology-the-need-for-public-regulation-and-corporate-responsibility/

[97]  J. Goldsmith, E. Burton. 2017. Why Teaching Ethics to AI Practitioners Is Important. *The AAAI-17 Workshop on AI, Ethics, and Society.*

[98]  Angela R. Bielefeldt, Madeline Polmear, Christopher Swan, Daniel Knight, Nathan Canney. 2017. An Overview of the Microethics and Macroethics Education of Computing Students in the United States. 2017 *IEEE Frontiers in Education Conference* (FIE).

[99]  https://ethicsunwrapped.utexas.edu/subject-area/behavioral-ethics

[100]  Shaul Shalvi, Francesca Gino, Rachel Barkan, Shahar Ayal. 2015. Self-Serving Justifications: Doing Wrong and Feeling Moral. *Current Directions in Psychological Science,* 24(2):125–130

[101]  The European Commission's draft ethics guidelines for trustworthy AI https://ec.europa.eu/digital-single-market/en/news/ethics-guidelines-trustworthy-ai

[102]  E. Mackintosh. 2019. Finland is winning the war of fake news. What it's learned may be crucial to Western democracy. CNN (special report).  https://edition.cnn.com/interactive/2019/05/europe/finland-fake-news-intl/

[103]  Omer Tene, Jules Polonetsky, Ahmad-Reza Sadeghi, 2018. Five Freedoms for the Homo Deus. *IEEE Computer and Reliability Societies.*

[104]  Teaching AI Ethics (TAIE) https://www.weforum.org/projects/teaching-ai-ethics

[105]  Artificial Intelligence at Google: Our Principles https://ai.google/principles/

[106]  Bruce M. McLaren. 2003. Extensionally defining principles and cases in ethics: An AI model. *Artificial Intelligence,* 150(1-2):145–181.

[107]  B.C. Stahl, and M. Coeckelbergh. 2016. Ethics of healthcare robotics: Towards responsible research and innovation. Robotics and Autonomous Systems. 86(2016)152-161. http://dx.doi.org/10.1016/j.robot.2016.08.018

[108]  CNBC. 2018. Amazon scraps a secret A.I. recruiting tool that showed bias against women. CNBC, 10 Oct 2018. *https://www.cnbc.com/2018/10/10/amazon-scraps-a-secret-ai-recruiting-tool-that-showed-bias-against-women.html*

[109]  Li Zhou. 2018. Is Your Software Racist? POLITICO https://www.politico.com/agenda/story/2018/02/07/algorithmic-bias-software-recommendations-000631 (Feb. 7, 2018)

[110]  John Braithwaite. 2017. Types of responsiveness. In *Regulatory Theory: Foundations and applications.* Peter Drahos (ed.). Anu press.

[111]  Anat R. Admati. 2017. A Skeptical View of Financialized Corporate Governance. *Journal of Economic Perspectives.* 31(3):131-150, July 2017.

[112]  Tracy Francis and Fernanda Hoefel. 2018. 'True Gen': Generation Z and its implications for companies. Mckinsey&Company, Nov. 2018. https://www.mckinsey.com

[113]  C. Duhigg. 2012. How companies learn your secrets. NY Times Magazine. Feb. 16, 2012 [Online]. https://www.nytimes.com/2012/02/19/magazine/shopping-habits.html

[114]  Ray Kurzweil. 2016. The Singularity Is Near: When Humans Transcend Biology. Penguin Books.

[115]  Andrew Ng. 2016. What Artificial Intelligence Can and Can't Do Right Now. *Harvard Business Review Online.* https://hbr.org/2016/11/what-artificial-intelligence-can-and-cant-do-right-now Accessed 21.7.2019

[116]  Gunningham and Sinclair, Smart Regulation. In *Regulatory Theory: Foundations and applications,* edited by Peter Drahos. Anu press. http://dx.doi.org/10.22459/RT.02.2017